# Transportation $CO_2$ emissions stayed high despite recurrent COVID outbreaks

Yilong Wang[1,7], Zhu Deng[2,7], Philippe Ciais[3], Zhu Liu[2], Steven J. Davis[4], Pierre Gentine[5,6], Thomas Lauvaux[3], and Quansheng Ge[1]

[1]Key Laboratory of Land Surface Pattern and Simulation, Institute of Geographic Sciences and Natural Resources Research, Chinese Academy of Sciences, Beijing, China
[2]Department of Earth System Science, Tsinghua University, Beijing, China
[3]Laboratoire des Sciences du Climat et de l'Environnement, CEA-CNRS-UVSQ- Université Paris Saclay, 91191, Gif-sur-Yvette CEDEX, France
[4]Department of Earth System Science, University of California, Irvine, 3232 Croul Hall, Irvine, CA, USA
[5]Department of Earth and Environmental Engineering, Columbia University, New York, New York 10027, USA
[6]Earth Institute, Columbia University, New York, New York 10027, USA.
[7]These authors contributed equally: Yilong Wang, Zhu Deng

e-mail: philippe.ciais@lsce.ipsl.fr; zhuliu@tsinghua.edu.cn; sjdavis@uci.edu

**After steep drops and then rebounds in transportation-related $CO_2$ emissions over the first half of 2020, a second wave of COVID-19 this fall has caused further—but less substantial—emissions reductions. Here, we use near-real-time estimates of daily emissions to explore differences in human behavior and restriction policies over the course of 2020.**

New COVID-19 outbreaks across Europe and North America this Fall were met with very different government responses. While some countries and U.S. states reinstated strict mobility restrictions (e.g., stay-at-home measures in France, Austria, New York, and California), some maintained more moderate restrictions (e.g., Germany, Netherlands and Switzerland). During the second outbreaks, a larger fraction of workers could commute with masks, and governments could better trace the spread of COVID-19 and adapt their restrictions. Here we use near-real-time estimates of ground transportation-related $CO_2$ emissions[1] from vehicle geolocation data to show that, from the new lockdown measures taken in Europe, emissions dropped by only 5.2% in November compared to 34.5% in April. Our results reveal a decoupling of transportation emissions from both mobility restrictions and the prevalence of COVID-19 cases or deaths between the first and second wave of outbreaks in 2020.

**Waxing and waning emissions**

Although mobility restrictions and stay-at-home measures caused substantial decreases in $CO_2$ emissions from ground transportation during the first COVID-19 lockdowns in March and April[1,2], changes occurring now during the second wave of outbreaks have not yet been characterized. We calculated daily emissions from ground transportation until the 1st of December 2020, from a unique dataset of near-real-time index of daily traffic congestion from TomTom (https://www.tomtom.com/en_gb/traffic-index/) covering 416 cities worldwide, 58% of which (239) being located in Europe. Our estimates are sensitive to subtle changes in ground transportation and mobility activities, providing a useful tool to track the emissions through the year. The emission time series in Fig. 1 show small or no reductions in emissions during the second wave of COVID-19 cases across most of Europe this Fall[3], reflecting less drastic mobility restrictions in many places.

After the spring drastic lockdowns and the summer of 2020, transportation emissions recovered to pre-pandemic (the week between March 2$^{nd}$ and March 8$^{th}$) levels in most European countries, but they remained lower in the U.S. (Fig. 1g). Beginning at the end of October, as renewed lockdowns went into effect in Europe, emissions dropped again. For example, France closed many workplaces and imposed stay-at-home requirements[4] beginning on October 30$^{th}$, leading to a 17.6% decrease in transportation emissions in November (compared to the pre-pandemic level). Less stringent restrictions caused no or smaller emissions reductions in Germany (0%), the U.K. (-9.1%), Poland (-4.3%) and Czechia (-4.7%). Although daily cases in the U.S. have surged to nearly 240,000 per day, no national restrictions have yet been implemented though some U.S. states and counties have announced new curfews and school closings since mid-November. U.S. transportation emissions have remained below the pre-pandemic level since mid-March, by 28.1% in November. In India, emissions have rebounded after the spring outbreak but remained -7.3% below pre-pandemic levels in November, when confirmed cases were still ~40% of the peak level in September. In China, emissions gradually recovered since May and have been at pre-pandemic (the week between January 13$^{rd}$ and January 19$^{th}$) level (-1.5%) since August, with few new COVID-19 cases.

Comparing the first and second waves of COVID outbreaks, transportation emissions in Europe and the U.S. dropped by 34.5%, and 51.8%, respectively, in April, and by 5.2% and 28.1% in November. Transportation emissions in China dropped by 55% in February. Globally, these changes of transportation emissions from different regions translate to an 8.8% reduction in February of 2020, largely in China, a 40.8% reduction in April, mainly in Europe and the U.S., and again an 18.8% reduction in November, mainly due to new and smaller decreases in Europe and a persistent drop in U.S.

**Relationship of emission reductions to restrictions and COVID-19 cases**

During the second lockdown, transportation emissions declined in Europe, but less so than during the first lockdown. This is partly due to somewhat less stringent restrictions (Fig. 2), but importantly, also due to the fact that people adapted their behavior to restrictions compared to the first wave. Fig. 2 shows the relationship between the emission reduction and the stringency indexes across European countries in the two waves. In the first wave, the emission reductions were more prominent in countries imposing stricter restrictions. However, during the second wave, the emission reductions is not tightly related to the severity of the restrictions. We also investigated the temporal sensitivity of emission reduction to the stringency index within each country during the two waves. The distribution of these temporal sensitivity across countries changed dramatically during the two waves (Fig. S1). Within a given country, the sensitivity of the emission reduction to the stringency index during the second wave (red bars) is significantly smaller than that during the first wave (blue bars). The dampened sensitivity of emission reductions to the stringency index within a country and across countries all implies that there is a decoupling of transportation emissions from government restrictions during the second wave. Meanwhile, we also found that changes in emissions were not related to pandemic severity during the second wave, as indicated by the lack of strong relationship between the emission reduction and the number of daily deaths (Supplementary Fig. S2a) or daily cases (Supplementary Fig. S2b) prior to announcing "lockdowns". The onset of emission reductions in fact occurred the day when policies restricting behavior were imposed by governments during the second wave, and not before, implying no significant voluntary changes in behavior in front of the increasing severity of outbreaks.

**Weekday and weekend differences**

In most countries, emissions on weekdays and weekends recovered at the same pace after the lockdown (the differences in recovered percentages between weekdays and weekends are less than 4%), as the restrictions were lifted after the end of the first lock-downs. However, a closer inspection at the data shown in Fig. 1 indicates that in China, emissions from weekdays recovered faster than those from weekends. We investigated relative emission changes for the most impacted week (with largest emission reduction) during the lockdown, the week before the release of first lockdown, and the recovery week (the third week) after the release of first lockdown, all compared to the pre-lockdown levels. We found that weekend emissions systematically recovered more slowly than weekday emissions in Czechia, China, India, Australia, Peru and Mexico (Fig. S3a). However, the lagging recovering of weekend emissions in these countries (colored lines in Fig. S3b), does not correspond to any different restriction stringency indexes than other countries (black triangles in Fig. S3b) that might disproportionately affect weekdays (e.g., the closing of schools and workplaces) and weekends (e.g. stay-at-home requirements). The slower recovery of weekends' emissions thus suggests changes in national behavior and/or closure of weekend tourist attractions. For instance, emissions in France decreased less during weekdays of the second wave than during weekends, as people seem to have opted to stay at home. Meanwhile, emissions in Germany and the U.K. have remained nearly unchanged during weekdays or the second wave and showed only a small reduction during weekends.

In summary, the decoupling of transportation emissions reductions from either the level of government restrictions or the prevalence of deaths and cases in a country during the second wave of the COVID-19 pandemic this fall reveals that confinement indexes and pandemic severity are not reliable predictors for estimating emission changes[2], and more robust estimates require daily activity data like those used in this study. Meanwhile, daily activity data identify subtle differences in behavior during the pandemics, such as the weekday-weekend distinct trajectories after lockdowns. The adaptation of traffic $CO_2$ emissions to COVID outbreaks lead us to predict that European ground transportation emissions will return to normal levels in 2021, with little legacy effects of the pandemic. In the U.S., by contrast, it is too early to conclude how the ongoing second outbreak will further affect $CO_2$ emissions from the ground transportation sector, but we can and will monitor those effects in near-real-time.

**Methods**

**TomTom congestion level.** TomTom congestion level data for 416 cities across 6 continents and 57 countries were collected from TomTom website (https://www.tomtom.com/en_gb/traffic-index/). The congestion level represents the extra time spent on a trip, in percentage, compared to uncongested conditions. TomTom congestion level data are available at a temporal resolution of one hour to 15 min. More than half of the TomTom cities are in Europe (Russia not included) and North America (93 cities).

**The estimation of $CO_2$ emissions from the ground transportation sector.** A sigmoid function was used to describe the relationship between the time series of daily mean TomTom index and actual car counts[1]. The daily relative magnitude of emissions was calculated for individual cities through 2019 and 2020, assuming that the relative magnitude of daily emissions was proportional to the relative magnitude of daily mean car counts. The time series were then weighted by city emissions to aggregate to the national mean. The national daily emissions were scaled such that the total annual emission in 2019 matches the estimation from ref[5].

**Lockdown severity.** The data from the Oxford Covid-19 Government Response Tracker[4] (OxCGRT) were used to describe the policy stringency. OxCGRT contains 19 indicators, grouped into "containment and closure policies", "economic policies", "health system policies" and "miscellaneous policies". Specific to the ground transportation, we selected three indicators from "containment and closure policies", namely "school closing",

"workplace closing" and "stay-at-home requirements". An indicator *j* at the time *t* is recorded on an ordinal scale ($v_{j,t}$) from 0 (no measures) to 3 (the strictest measures). Each of these indicators contains a flag ($f_{j,t}$) to tell whether the policy is taken for a targeted geographical region ($f_{j,t}=0$) or nationwide ($f_{j,t}=1$).

The stringency of an individual indicator is calculated as:

$$I_{j,t} = 100 \frac{v_{j,t} - 0.5(f_{j,t}-1)}{3} \quad (1)$$

The scores of the three indicators are simply averaged to get the overall stringency index:

$$\text{index} = \frac{1}{3} \sum_{j=1}^{3} I_{j,t} \quad (2)$$

For a large country like China, the strict policy for a targeted region may indicate an overall high level of restrictions for the country. For example, the strictest restrictions were taken in Beijing in June when new cases were reported, while the other provinces did not strengthen their restrictions. According to Eq. 1, this will result in a score of $I_t=83.3$ for a given indicator and a high value in the overall stringency index.

**Pandemic severity.** The data of daily new cases and daily new deaths from the Johns Hopkins University Center for Systems Science and Engineering (https://coronavirus.jhu.edu/data)[3] were used to indicate the severity of the pandemic of COVID-19. The daily new cases and daily new deaths were smoothed with a seven-day rolling window.

**Lockdown periods.** The lockdown periods for the countries impacted by the COVID-19 outbreaks were collected from https://en.wikipedia.org/wiki/COVID-19_pandemic_lockdowns, which are defined as the periods when non-pharmaceutical anti-pandemic measures are taken including: closing of schools and kindergartens, closing of non-essential shops (shops and stores apart from food, doctors and drug stores), closing of non-essential production, cancellation of recreational venues and closing of public places, curfews, stay-at-home orders and total movement control.

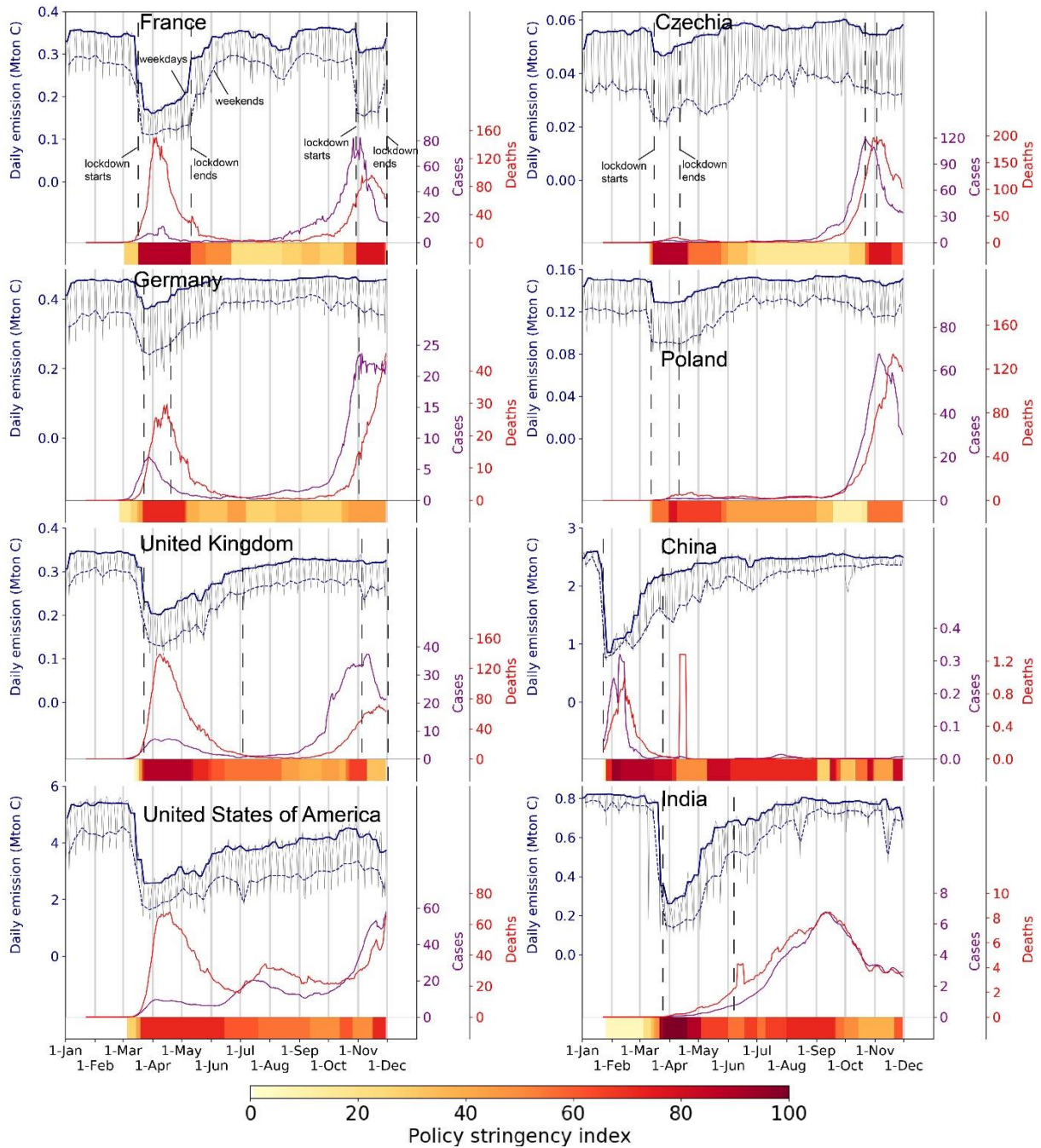

**Fig. 1. $CO_2$ emissions from the ground transportation sector in 2020.** Thin grey lines are the time series of daily emissions. Thick solid dark blue lines connect mean emissions on weekdays (with festivals excluded), and thin solid dark blue lines connect mean emissions on Saturday and Sunday (with festivals excluded). The purple and red lines show the seven-day moving average of daily new cases per 10,000 people and daily new deaths per 1,000,000 people. The color patches under each plot represent the policy stringency indexes, according to the Oxford Covid-19 Government Response Tracker (see Methods). Thick solid vertical grey lines separate different months, and thin dashed vertical black lines depict the lockdown periods (see Methods).

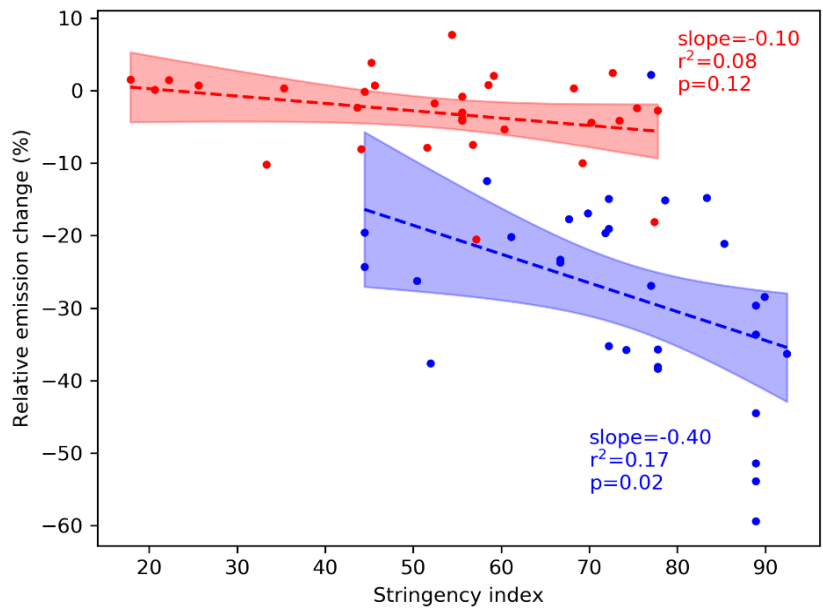

**Fig 2. Relationship between emission reductions and stringency indexes of restrictions among the two waves of COVID-19.** The emission reductions and stringency indexes are plotted for the two periods between April 1st and April 28th (blue) and between November 1st and November 28th (red). Each dot represents the relationship for a given country. The dashed lines are the linear regressions with shaded area being the 95% confidence intervals.

**Acknowledgement**

This study was financially supported by Strategic Priority Research Program of Chinese Academy of Sciences "Science and Technology Project of Beautiful China Ecological Civilization Construction" (No. XDA23100400), the National Natural Science Foundation of China (42001104) and National Key Research and Development Program of China (2017YFA0605303).

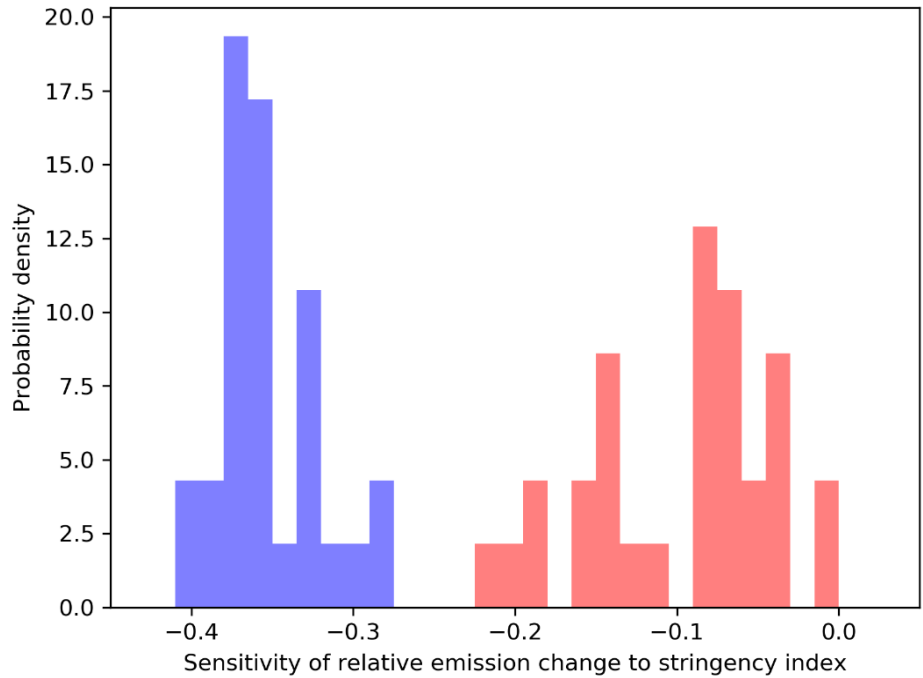

**Fig. S1. Distribution of temporal sensitivity of emission reductions to stringency indexes during the two waves.** For each country, an 8-week time window covering the pre-lockdown and lockdown periods is used to compute the sensitivity. The histograms represent the distribution of these sensitivities across countries. The time window from March $1^{st}$ to April $25^{th}$ is used for the first wave, while the time window from October $1^{st}$ to November $25^{th}$ is used for the second wave.

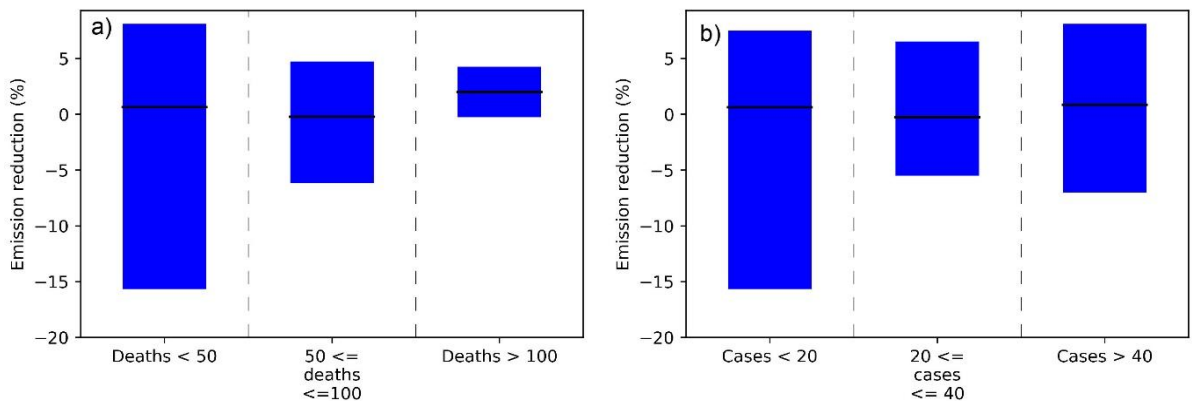

**Fig. S2. Emission reductions and the pandemic severity in terms of daily new deaths per 1,000,000 people (a) and daily new cases per 10,000 people (b).** For countries announcing the second lockdowns, the emission reductions and pandemic severity are calculated as the mean of the

week prior to lockdowns. For countries not announcing the second lockdowns, the emission reductions and pandemic severity is calculated are calculated as the mean of the week from October 25$^{th}$ to October 31$^{st}$.

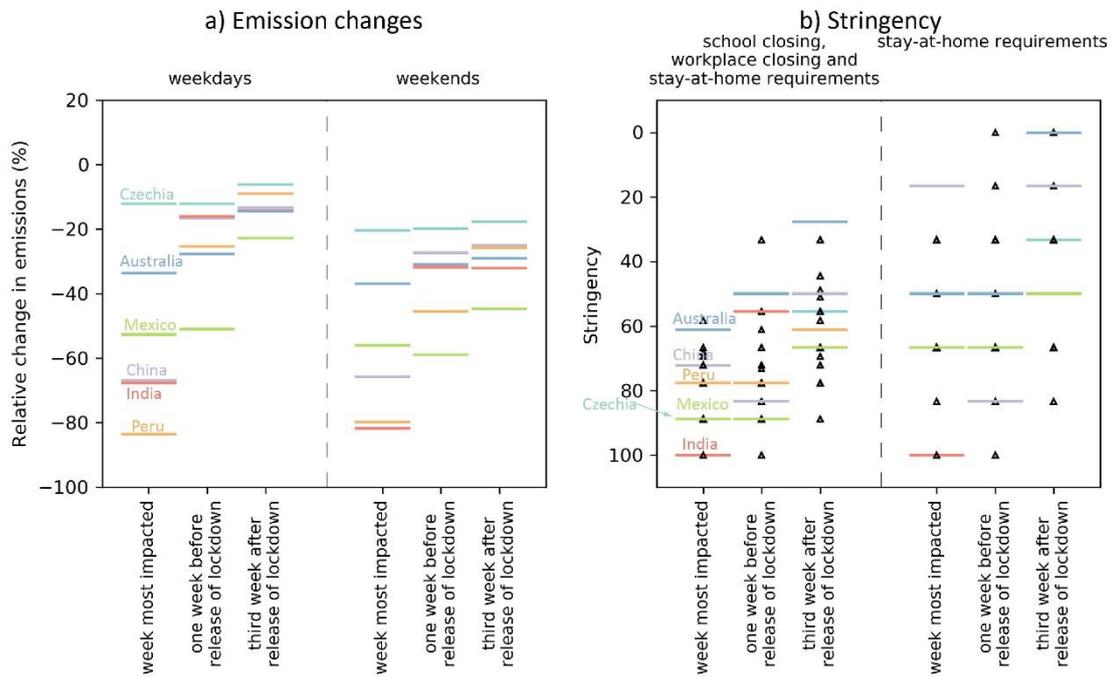

**Fig. S3. Recovery of emissions for weekdays and weekends after lockdown.** The black triangles represent the policy stringencies for other countries where no slow recovery of weekends' emissions are found. The emission changes are defined as relative changes compared to the mean pre-COVID emissions, which are between Jan 13$^{rd}$ and Jan 19$^{th}$ for China and between Mar 2$^{nd}$ and Mar 8$^{th}$ for other countries.